# 3D-printed facet-attached optical elements for connecting VCSEL and photodiodes to fiber arrays and multi-core fibers

Pascal Maier,[1,2,†,*] Yilin Xu,[1,2,†] Mareike Trappen,[1,2] Matthias Lauermann,[3] Alexandra Henniger-Ludwig,[4] Hermann Kapim,[4] Torben Kind,[5] Philipp-Immanuel Dietrich,[1,2,3] Achim Weber,[3,5] Matthias Blaicher,[1,2] Clemens Wurster,[6] Sebastian Randel,[1] Wolfgang Freude,[1] and Christian Koos[1,2,3,*]

[1]*Institute of Photonics and Quantum Electronics (IPQ), Karlsruhe Institute of Technology (KIT), Engesserstr. 5, 76131 Karlsruhe, Germany*
[2]*Institute of Microstructure Technology (IMT), Karlsruhe Institute of Technology (KIT), Hermann-von-Helmholtz-Platz 1, 76344 Eggenstein-Leopoldshafen, Germany*
[3]*Vanguard Automation GmbH, Gablonzer Str. 10, 76185 Karlsruhe, Germany*
[4]*Rosenberger Hochfrequenztechnik GmbH & Co. KG, Hauptstr. 1, 83413 Fridolfing, Germany*
[5]*ficonTEC Service GmbH, Im Finigen 3, 28832 Achim, Germany*
[6]*Rosenberger OSI GmbH & Co. OHG, Endorferstr. 6, 86167 Augsburg, Germany*
[†]*both authors contributed equally to this work.*
* *pascal.maier@kit.edu; christian.koos@kit.edu*

**Abstract:** Multicore optical fibers and ribbons based on fiber arrays allow for massively parallel transmission of signals via spatially separated channels, thereby offering attractive bandwidth scaling with linearly increasing technical effort. However, low-loss coupling of light between fiber arrays or multicore fibers and standard linear arrays of vertical-cavity surface-emitting lasers (VCSEL) or photodiodes (PD) still represents a challenge. In this paper, we demonstrate that 3D-printed facet-attached microlenses (FaML) offer an attractive path for connecting multimode fiber arrays as well as individual cores of multimode multicore fibers to standard arrays of VCSEL or PD. The freeform coupling elements are printed *in situ* with high precision on the device and fiber facets by high-resolution multi-photon lithography. We demonstrate coupling losses down to 0.35 dB along with lateral 1 dB alignment tolerances in excess of 10 μm, allowing to leverage fast passive assembly techniques that rely on industry-standard machine vision. To the best of our knowledge, our experiments represent the first demonstration of a coupling interface that connects individual cores of a multicore fiber to VCSEL or PD arranged in a standard linear array without the need for additional fiber-based or waveguide-based fan-out structures. Using this approach, we build a $3 \times 25$ Gbit/s transceiver assembly which fits into a small form-factor pluggable module and which fulfills many performance metrics specified in the IEEE 802.3 standard.

## 1. Introduction

In recent years, increasing throughput of optical communication systems has primarily relied on higher symbol rates and advanced modulation formats [1]. Giving the steadily rising demand for communication bandwidth, however, the limits of these approaches are becoming increasingly obvious [2,3], especially for short-reach intra-datacenter links, where cost- and energy-efficient implementation is key. Parallel transmission via spatially separated channels is seen as an attractive alternative for bandwidth scaling with linearly increasing technical effort [3,4]. To maintain the associated fiber installations manageable, significant effort has been spent to replace single fibers by more compact fiber ribbons [5−13], comprising fiber arrays (FA), or by multicore fibers (MCF) [14−23]. However, low-loss coupling of light between FA or MCF and standard linear arrays of vertical-cavity surface-emitting lasers (VCSEL) or photodiodes (PD) still remains challenging. Current solutions for coupling of VCSEL or PD to FA rely on, e.g., high-precision injection-molded plastic parts that contain refractive and reflective optical elements to adapt the spot size of the emitted light for efficient coupling to the corresponding fiber [5−7,10,12,24]. However, these schemes often require multi-step assembly processes, starting with a precise mechanical socket that needs to be carefully aligned and fixed to the on-board VCSEL / PD array. This socket is equipped with a pluggable interface to a detachable mechanical connector that has to be glued to a FA in a separate step and that can then be connected to the on-board socket. These schemes require high-precision visual alignment of the socket with respect to the VCSEL / PD array as well as precise mounting of the connector to the FA, while typical losses range between 1 dB and 2 dB [10,25]. Moreover, these schemes are not applicable to MCF, which either have to rely on fan-out structures to address individual fiber cores [14,26,27], often in conjunction with custom connector and fiber arrangements [14−16,28], or which require device [17−21,23] or grating coupler (GC) [22] arrays in non-standard 2D arrangements that are precisely matched to the cross-section of the respective MCF. Such solutions are technically complex and challenging to scale, in particular when it comes to compact short-reach data-center transceivers that are subject to stringent constraints in footprint and in assembly costs.

In this paper, we show that 3D-printed facet-attached microlenses (FaML) [29] may offer an attractive alternative for efficiently connecting multimode fiber arrays (MM-FA) as well as individual cores of multimode multicore fibers (MM-MCF) to standard arrays of VCSEL or PD having industry-standard pitches of, e.g., 250 µm. The FaML are printed directly on the device and fiber facets by multi-photon lithography [30], thereby ensuring sub-100 nm precision in a fully automated fabrication step. The freeform coupling elements are designed to collimate the associated beams to large diameters of tens of micrometers, which greatly relaxes alignment tolerances such that subsequent assembly steps can entirely rely on passive positioning using industry-standard machine vision. We demonstrate the viability of the proposed concepts in a series of proof-of-concept experiments. In a first set of experiments, we show connections between VCSEL / PD arrays and MM-FA, achieving average coupling losses as low as 0.35 dB for the transmitter (Tx) and 0.70 dB for the receiver (Rx) along with lateral 1 dB alignment tolerances of ± 17 µm (Tx) and ± 62 µm (Rx), respectively. To the best of our knowledge, these results are among the lowest losses and the highest alignment tolerances so far demonstrated for coupling between VCSEL / PD arrays and MM-FA. In a second set of experiments, we extend this concept to MM-MCF containing densely spaced cores with a separation of 39 µm. Using appropriately designed FaML, these cores can be connected to VCSEL and PD that are arranged in linear arrays with a standard pitch of 250 µm, reaching average coupling losses of 0.67 dB (Tx) and 0.63 dB (Rx) along with lateral 1 dB alignment tolerances of ± 18 µm (Tx) and ± 25 µm (Rx), respectively. To the best of our knowledge, these experiments represent the first demonstration of a coupling interface that connects individual cores of an MCF to VCSEL / PD arranged in a standard linear array without the need for additional fiber-based or waveguide-based fan-out structures. Using this approach, we finally build a 3 × 25 Gbit/s transceiver assembly which fits into a small form-factor pluggable (SFP) module and which fulfills many performance metrics specified in the IEEE 802.3 standard.

## 2. Coupling concept

Figure 1 illustrates the concept of using 3D-printed facet-attached microlenses (FaML) for connecting linear arrays of vertical-cavity surface-emitting lasers (VCSEL) and photodiodes (PD) to fiber arrays (FA) and multicore fibers (MCF). The VCSEL and PD arrays transmit light to or receive light from the associated FA or MCF, which are glued into an industry-standard mechanical transfer (MT) ferrule. MT ferrules are commercially available at high quality and low cost and can be handled with standard gripper tools. Moreover, in future implementations, MT ferrules could allow to arrange fibers in two-dimensional arrays, Fig. 1(a), thereby increasing the number of parallel fiber channels.

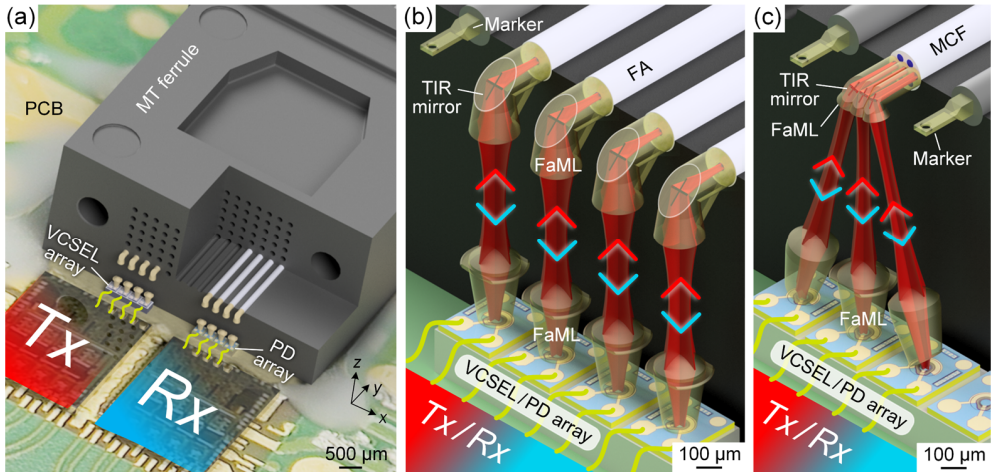

**Fig. 1.** Concept of a multi-lane transceiver assembly using 3D-printed facet-attached microlenses (FaML) for connecting linear arrays of vertical-cavity surface-emitting lasers (VCSEL) and photodiodes (PD) to fiber arrays (FA) or multicore fibers (MCF). The VCSEL and PD arrays transmit light to or receive light from the associated FA or MCF, which are glued into an industry-standard mechanical transfer (MT) ferrule. The FaML are designed to redirect, expand, and collimate the beams emitted or accepted by the VCSEL, the PD, the FA or the MCF such that the alignment tolerances increase both in axial and transverse direction with respect to the beam. **(a)** Overview of a transceiver assembly. Transmitter (Tx), receiver (Rx), and MT ferrule are mounted to a printed circuit board (PCB) in a small form-factor pluggable (SFP) layout. **(b)** Four linearly arranged VCSEL or PD of a Tx or Rx array, respectively, coupled to four fibers of a FA. The device and fiber facets are equipped with 3D-printed FaML, which contain curved refractive surfaces as well as total-internal-reflection (TIR) mirrors to shape and redirect the beams. 3D-printed markers facilitate passive alignment during the assembly process. **(c)** A linear array of three VCSEL or PD is used to couple to three cores of a single MCF. The receiving (emitting) MCF are glued to the MT ferrule such that the axes of three fiber cores lie in a common z-normal plane. To compensate for the pitch mismatch between the MCF cores and the VCSEL / PD, the outer FaML are designed to produce beams that are slightly inclined with respect to the z-direction. Note that this concept could in principle be extended further to also utilize the remaining cores of the depicted seven-core MCF. This could open a path to using two-dimensional device arrays in combination with one-dimensional MCF arrays and more complicated FaML arrangements on the fiber facets.

The assemblies illustrated in Fig. 1 can be build up in two steps: In a first step, FaML are printed to the VCSEL / PD chips and to the fiber facets using multi-photon lithography [30]. This approach allows to use the full design freedom of 3D-printed freeform structures and ensures highly precise alignment with deviations well below 100 nm in a fully automated fabrication step. The FaML-equipped devices can then be combined in a separate assembly step using passive alignment techniques based on industry-standard camera-based machine vision and height measurements. To this end, the FaML illustrated in Fig. 1 are designed to redirect, expand, and collimate the beams emitted or accepted by the VCSEL, the PD, the FA or the MCF such that the alignment tolerances increase both in axial and in transverse direction of the beam. The simultaneously decreased angular tolerance can be usually accepted when using industry-standard assembly machinery. Note that the FaML approach can also be used for increasing the alignment tolerances of single-mode coupling interfaces [29]. Note also that, for assemblies like those shown in Fig. 1, coupling through FaML may have distinct advantages over the photonic wire bonding schemes used for fiber-chip interfaces in previous demonstrations [31−33]. Specifically, FaML-based coupling schemes allow to bridge comparatively large distances in the millimeter range, possibly with intermediate free-space micro-optical elements. This is impossible with photonic wire bonds (PWB), for which both facets have to be accessible within a single write field (typical size 300 µm × 300 µm) and within the limitations of a rather small working distance of the high-NA lithography objective (typically 250 µm). Moreover, FaML can be printed to device and fiber facets prior to module assembly. It is thus not necessary to expose the full assembly to the solvents used in the development process, which is unavoidable for PWB-based assembly workflows.

Figure 1(b) illustrates the concept for connecting individual VCSEL or PD of a linear Tx or Rx array, respectively, to a MM-FA having the same pitch. On the VCSEL (PD) side, a single FaML is used to emit (receive) a collimated beam. Within the FaML on the fiber facet, the beam is redirected by a total-internal-reflection (TIR) mirror. We further demonstrate that VCSEL and PD in a linearly arranged array can be connected to individual cores of a single MM-MCF using appropriately designed FaML, see Fig. 1(c). In the example shown in Fig. 1(c), the MCF are glued to the MT ferrule in a well-defined orientation such that the axes of three fiber cores lie in a common $z$-normal plane. To compensate for the pitch mismatch between the MCF cores and the VCSEL / PD, the outer FaML are tilted such that the emitted or received beams are slightly inclined with respect to the $z$-direction. We have performed proof-of-concept experiments of both the arrangements shown in Fig. 1(b) and (c), which we describe in more detail in the following sections. Note that the concept shown in Fig. 1(c) could in principle be extended further to also utilize the remaining cores of the depicted seven-core MCF. This could open a path to using two-dimensional device arrays in combination with one-dimensional MCF arrays and more complicated FaML arrangements on the fiber facets.

## 3. Connecting VCSEL / PD to multimode fiber arrays (MM-FA)

We demonstrate the viability of the coupling concept shown in Fig. 1(b) by connecting four linearly arranged VCSEL (Broadcom AFCD-V64JZ, $\lambda = 850$ nm [34]) and PD (Broadcom SPD2025-4X, responsivity $S = 0.5$ A/W [35]) of a pair of Tx / Rx modules to a MM-FA, see Fig. 2(a). The MM-FA feature a pitch of 250 µm, matched to the pitch of the VCSEL / PD array, and are equipped with MM fibers having a core diameter of $2a = 26 \mu$m. Figure 2(b) provides a more detailed view of the Tx VCSEL array connected to four MM fibers of the MT ferrule. The Rx PD array is connected via an identical arrangement but differently designed FaML. Inset (i) of Fig. 2(b) shows the FaML on the fiber side, each comprising a TIR mirror and a beam-expander lens, while Inset (ii) depicts the expander FaML at the VCSEL. Technical drawings with further details are given in Fig. 2(c), where the left-hand side shows a projection of a FaML pair along the $y$-direction ("front view"). For better visibility, the distance between the two FaML has been reduced in the drawing. Note that the MT ferrule used in our experiment was equipped with MM fibers (OFS MCF-MM-7-39) containing seven cores each, out of which only the central one was used, see cross section at the upper left in Fig. 2(c). The right-hand side of Fig. 2(c) shows a cut through the FaML arrangement along the $x$-normal plane through the line A–A' as indicated on the left.

To achieve high coupling efficiency, the shapes of the refractive surfaces are optimized using a home-made wave-propagation algorithm, which is based on the theory described in [36] and which has been successfully used for similar tasks [37,38]. Multimode light propagation through the FaML assembly is emulated by using Gaussian beams with an effective wavelength $\lambda_{\text{eff}} = M^2 \lambda$, where $M^2 = 4.2 \ldots 5.4$ is the measured beam quality factor, see Appendix A and B for a more detailed explanation. At the Tx, the refractive Surface S1 on the laser side is designed to produce a mode-field diameter (MFD) of 49 µm half-way between the two FaML. The beam then enters the FaML on the Tx fiber facet through refractive Surface S2, which is designed to illuminate the central core of the MM-MCF up to 70 % of the core radius $a = 13 \mu$m upon redirection by a flat TIR mirror with Surface S3, see Section 3.2 for a more detailed discussion of the launch conditions [39]. The distance between the apices of Surfaces S1 and S2 amounts to $d = 1150 \mu$m, Fig. 2(c). We also connected a PD array to a MM-FA. At the Rx side, the refractive Surface S2 is designed to produce a MFD of 61 µm half-way between the two FaML. The refractive Surface S1 on the PD is designed to focus the incoming beam to a spot with a MFD of 12 µm, well within the light-sensitive PD area, which has a diameter of 32 µm. As before, the distance between the apices of Surfaces S1 and S2 amounts to $d = 1150 \mu$m.

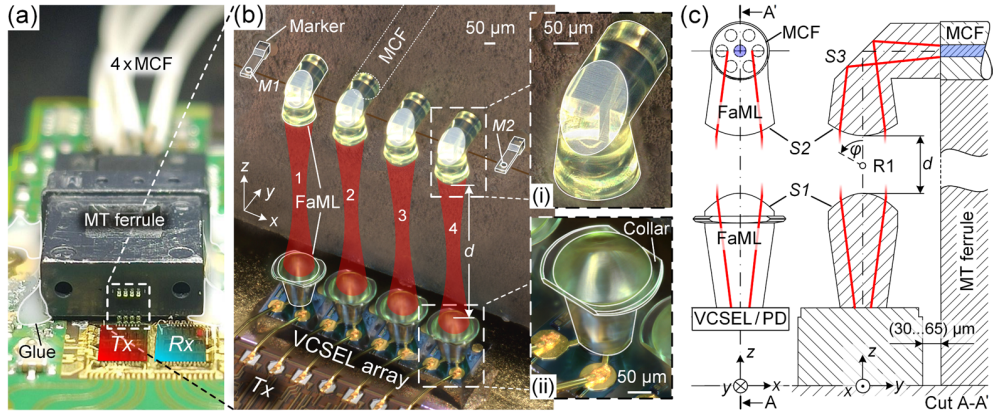

**Fig. 2.** Photograph and technical drawings of a four-lane transceiver (Channels 1, 2, 3, and 4), implemented according to the concept in Fig. 1(b). Four VCSEL are connected to the cores of four MM fibers, arranged in an array with a standard 250 µm pitch. **(a)** Overview photograph of the Tx side. **(b)** Microscope image showing the Tx VCSEL array and the associated FaML. The PD array is connected with a similar arrangement. Inset (i) shows the FaML on the fiber side, containing TIR mirrors and beam-expander lenses. Inset (ii) depicts the FaML on the VCSEL facet. A collar surrounds the VCSEL FaML for facilitating image recognition during passive assembly. Two markers are 3D-printed to the facet of the MT ferrule for reliable detection of the $z$-position and tilt correction. **(c)** Technical drawings of coupling structures. Left: Projection of a FaML pair along the $y$-direction ("front view"). Right: Cross-sectional view along the $x$-normal plane through the line A–A' shown on the left. The beam is shaped and redirected by the refractive Surfaces S1 and S2 as well as by the TIR mirror S3. The distance between the apices of Surfaces S1 and S2 amounts to $d = 1150\,\mu m$. For measuring the angular misalignment loss, the MT ferrule is rotated in $\varphi$-direction about the rotation axis R1, that is parallel to the $x$-axis and that passes through the mid-points between the corresponding FaML apices.

### 3.1 Module assembly

As a first step of the assembly process, the VCSEL array and the PD array are mounted to the PCB next to the Tx and Rx driver IC. The FaML are then printed separately to the VCSEL / PD chips and to the facet of the MT ferrule using high-resolution multi-photon lithography [30]. Note that this sequence is not mandatory – the FaML can also be printed to the device facets on a wafer level, before mounting the devices to the PCB. The two printing steps are carried out using negative-tone photoresists with a refractive index $n = 1.54$ at 850 nm. Precise alignment of the FaML relative to the respective facet is ensured by machine vision. After exposure, the fabricated structures are developed in propylene-glycol-methyl-etheracetate (PGMEA) for 15 minutes, flushed with isopropanol, and subsequently blow-dried. For more details on the fabrication technique, see Appendix C. Note that the FaML on the VCSEL / PD are fabricated from a photoresist (VanCore B, Vanguard Automation GmbH), which is compatible with industry-standard reflow soldering processes and for which long-term stability has been experimentally confirmed in damp-heat tests at a temperature of 85 °C and at a relative humidity of 85 %, see Appendix D.

A custom pick-and-place machine is then used to mount the MT ferrule to the PCB in a fully automated process, relying solely on industry-standard camera-based machine vision and height measurements with a confocal chromatic imaging sensor (Precitec CHRocodile S [40]). During the assembly procedure and the associated measurements, the MT ferrule is gripped by an air-pressure activated tool (gripper), which features mechanical stops that keep the outer edges of the MT ferrule aligned. In addition, a cable management system included in our pick-and-place machine ensures that the fiber strands attached to the MT ferrule do not experience any strain. During the assembly process, the PCB is fixed to the assembly zone of the pick-and-place machine using a holder placed on a vacuum chuck. For alignment, we first detect the centers of the FaML belonging to the outermost VCSEL / PD channels and extract the connecting line. For improving the accuracy of the image recognition, the FaML on the VCSEL / PD are surrounded by a collar, see Fig. 2(b) and lower inset. The MT ferrule is then gripped by the air-pressure activated tool, and the line between the two 3D-printed marker holes, denoted M1 and M2 in Fig. 2(b), is extracted. In a next step, the MT ferrule is moved in *x*- and *y*-direction to align the connection M1–M2 to the formerly found connecting line defined by the VCSEL / PD FaML. The ferrule is then laterally shifted along the M1-M2 connection for positioning the upper FaML exactly vertically above the corresponding lower FaML. In doing so, we iteratively correct for any tilt of the fiber plane in relation to the plane on which the VCSEL / PD chips are mounted, until an angular tilt of the two planes by less than 0.1° is reached. Finally, the *z*-position of the MT ferrule is adjusted to the designed distance of $d = 1150\,\mu\text{m}$ between the FaML apices using the chromatic confocal imaging sensor. This final position is fixed by applying a low-shrinkage UV-curable epoxy glue to the four corners of the MT ferrule (EMI Optocast 3410 Gen2), Fig. 2(a). Until the glue is cured, we maintain the position of the MT ferrule by electronically stabilizing the height of the stage.

*3.2 Alignment tolerance and coupling loss*

Prior to applying the glue, we quantify the alignment tolerances by moving the MT ferrule in *x*- and *y*-direction or by introducing a known tilt angle *φ*. The tilt is defined with respect to a rotation axis R1 that is parallel to the *x*-axis and that passes through the mid-points between the corresponding FaML apices, see Fig. 2(c). Note that due to the strongly expanded spot size of the beams, the Rayleigh length is rather large (approx. 2 mm), such that misalignment along the beam direction (*z*-direction) does not impair the coupling efficiency to a significant degree. We confirmed this notion by repeating the alignment process 100 times and by comparing the positions from the passive and active alignment procedures [41]. We found that the *z*-position can be reproduced with a standard deviation of only 1.5 µm and a maximum error of 7 µm. The corresponding excess coupling loss amounts to less than 0.1 dB and can be safely neglected.

For measuring the coupling losses, we either let the VCSEL emit light into the central core of one of the four Tx MM-MCF, or we receive light from one of the MM-MCF cores by the associated PD. In the experiment, we measure the misalignment excess loss by comparison to loss measured in the optimum position. Note that the various modes of an MM fiber may experience vastly different propagation losses and that a reliable quantification of coupling losses should hence refer to a "steady-state" modal distribution that is reached in the limit of long propagation distances [39,42]. Such a steady-state distribution of modal power can approximately be achieved by a so-called limited phase-space (LPS) launch [39]. In this approach, the fiber is fed by a multimode excitation field, having a Gaussian power distribution for which the $1/e^2$ diameter of the intensity profile corresponds to 70 % of the fiber-core diameter while the $1/e^2$ divergence angle is adjusted to 70 % of the maximum acceptance angle, which is found in the center of the core in case of graded-index MM fibers. In our experiment, it was not possible to simultaneously fulfill both requirements. In our lens design, we therefore adjusted the beam diameter to 70 % of the core diameter, while the acceptance cone was filled by more than 70 %. To arrive at a realistic estimate of the coupling losses, we used a cladding-mode stripper consisting of a piece of fiber that was coiled around an approximately 20 mm-thick metal rod [43]. The power of the resulting steady-state modal power distribution was finally measured by an integrating sphere.

For active alignment of the MM-FA to the Tx VCSEL array, we estimate absolute coupling losses of 0.05 dB, 0.25 dB, 0.62 dB, and 0.46 dB for Channels 1, 2, 3, and 4 as defined in Fig. 2(b), respectively. This leads to an average coupling loss of 0.35 dB with a standard deviation of 0.25 dB. Note that this standard deviation is very likely overestimated: For quantifying the losses at the Tx side, we first measured the output power of the VCSEL array to obtain a baseline to which we can refer the fiber-coupled power levels. During these pre-characterization experiments, it was unfortunately not possible to individually measure the output power of each VCSEL in the Tx array. This problem was caused by an electronics-related issue that prevented us from individually switching the four VCSEL in the Tx array on and off. Since the VCSEL are closely spaced, we could hence only capture the overall power emitted by all four VCSEL, from which we calculate the average emission power per device. The reported loss variations hence contain the variations of the VCSEL output powers within the array, which are specified as $\pm 7.5\,\%$ ($\pm 0.3$ dB) in the associated data sheet [34]. The true variation of the FaML coupling efficiency is hence presumably below the measured 0.25 dB.

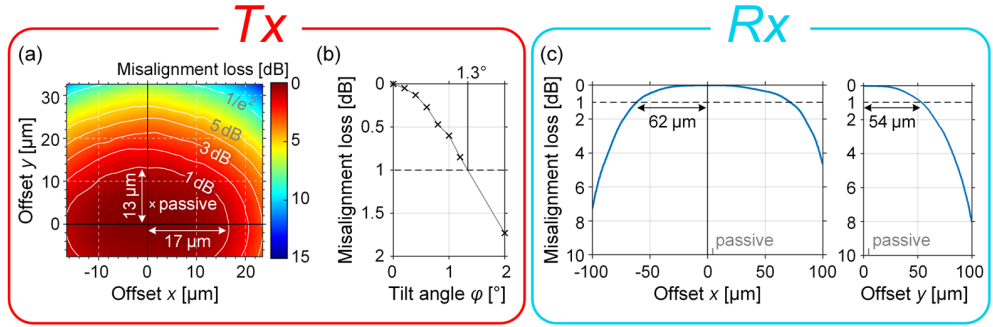

**Fig. 3.** Measured misalignment excess loss of a four-lane transmitter (Tx, VCSEL) and receiver (Rx, PD), where each device is coupled to a corresponding core of a MM-FA. We move the MT ferrule along the *x*- and *y*-direction, and we rotate it in *φ*-direction about the rotation axis R1, see Fig. 2(c). The excess loss is zero at the optimum position. Cladding-mode strippers are used to approximate a steady-state distribution of modal power within the MM fiber. **(a)** Excess loss for lateral misalignment measured for one of the four nominally identical FaML-based VCSEL-fiber connections at the Tx. White curves indicate the contour lines of constant loss. For a 1 dB loss, a deviation of $\Delta x = \pm 17\,\mu\text{m}$ in *x*-direction and of $\Delta y = \pm 13\,\mu\text{m}$ in *y*-direction can be tolerated. Using active alignment, the absolute coupling loss of the displayed VCSEL-fiber connection (Channel 1 in Fig. 2(b)) amounts to 0.05 dB, measured in the 0 dB point. The position obtained from automated passive assembly at $(x = 1\,\mu\text{m}, y = 5\,\mu\text{m})$ is indicated by a white cross ("passive"), leading to an excess loss of 0.15 dB. **(b)** Angular misalignment excess loss measured at the Tx by rotating the MT ferrule in *φ*-direction about axis R1 in Fig. 2(c). The angular 1 dB tolerance is 1.3°. **(c)** Lateral misalignment excess loss measured at the Rx for one of the four nominally identical FaML-based fiber-PD connections. For simplicity, we restrict our experiment to linear movements along the *x*-direction (left panel) and along the +*y*-direction (right panel). Note that the movement in –*y*-direction is restricted to avoid collision of the MT ferrule and the PD chip, see Fig. 2(c). For a 1 dB loss, a deviation of $\Delta x = \pm 62\,\mu\text{m}$ in *x*-direction and of $\Delta y = \pm 54\,\mu\text{m}$ in *y*-direction can be tolerated. The position obtained from automated passive assembly at $(x = 4\,\mu\text{m}, y = 4\,\mu\text{m})$ is indicated by grey tick labels ("passive"). The associated excess loss can be neglected. The absolute average coupling loss in the optimum position $(x = 0\,\mu\text{m}, y = 0\,\mu\text{m})$ amounts to 0.70 dB.

In a next step, we measure the lateral misalignment excess loss related to the optimum position $(x = 0\,\mu\text{m}, y = 0\,\mu\text{m})$ for one of the four nominally identical FaML-based VCSEL-fiber connections of the Tx module, see Fig. 3(a). The absolute coupling loss of the displayed VCSEL-fiber connection, which corresponds to Channel 1 in Fig. 2(b), amounts to the above-mentioned 0.05 dB, measured in the 0 dB point of Fig. 3(a). White curves indicate the contour lines of constant loss. For a 1 dB loss, a deviation of $\Delta x = \pm 17\,\mu\text{m}$ in *x*-direction and of $\Delta y = \pm 13\,\mu\text{m}$ in *y*-direction can be tolerated. The position obtained by the machine-vision-based automated passive assembly procedure is also indicated by a white cross in the first quadrant of Fig. 3(a) at $(x = 1\,\mu\text{m}, y = 5\,\mu\text{m})$, see white lettering "× passive". We attribute the larger positioning uncertainty along the *y*-direction to errors in measuring the tilt of the VCSEL bar about the *x*-axis. Note that the long side of the VCSEL bar is aligned along the *x*-direction, while the short axis is parallel to the *y*-direction, see Fig. 2(b). Since our tilt measurements rely on sampling the *z*-position of the chip surface with the confocal chromatic imaging sensor, we thus expect larger uncertainties for tilts about the long axis (*x*) as compared to tilts about the short axis (*y*). For an incorrectly compensated tilt about the *x*-axis, the emitted beam will hit the corresponding FaML on the fiber facet with a small offset in the *y*-direction, which is observed in our measurements shown in Fig. 3(a). Still, the automated passive alignment loss is well inside the 1 dB tolerance. To evaluate the angular alignment tolerance, we rotate the MT ferrule by the tilt angle *φ* about the rotation axis R1, see Fig. 2(c). Note that the movement in –*φ*-direction was restricted to avoid collision of the MT ferrule and the VCSEL chip, see Fig. 2(c). The results are plotted in Fig. 3(b). The angular 1 dB tolerance is 1.3°.

In a next step of the experiment, we use passive alignment techniques, relying on the top- and bottom-view cameras of our assembly system and on the confocal chromatic imaging sensor (Precitec CHRocodile S [40]). We obtain a 0.15 dB penalty for Channel 1 with respect to the actively aligned MM-FA, see Fig. 3(a), and we expect similar penalties for the other channels. This would lead to an average coupling loss of 0.50 dB for all four channels.

These measured coupling losses and alignment tolerances can well compete with those obtained for more complex multi-step assembly techniques relying on precision molded plastic parts which were actively aligned to the underlying VCSEL array [5−7,10,12]. For active alignment, lateral 1 dB alignment tolerances of $\pm 17$ µm in combination with minimum coupling losses of 0.5 dB have been demonstrated in [7]. In a similar experiment [12], the lateral 1 dB alignment tolerances were increased to $\pm 35$ µm at the expense of a slightly higher coupling loss of 1 dB. Note that the MM fibers used in our experiment have core diameters of only 26 µm – significantly smaller than the more standard core diameters of 50 µm, that have, e.g., been used in [12]. Using larger core diameters would further increase the alignment tolerances in our experiment. For passive alignment, our estimated average losses of 0.5 dB can well compete with those obtained in previous demonstrations of coupling interfaces between VCSEL and MM-FA [8], where losses down to 0.7 dB along with alignment tolerances of $\pm 18$ µm have been reached for comparatively large core diameters of 62.5 µm.

For measuring the coupling losses to the PD at the Rx side, we inject light into each of the central cores of the four Rx fibers and measure the power of the associated steady-state distribution after the mode stripper using an integrating sphere. The power incident on the PD is obtained from the respective photocurrents and the data-sheet specification of the responsivity $S$. For active alignment of the MM-FA to the Rx PD array, we measure absolute coupling losses of 0.63 dB, 0.66 dB, 0.77 dB, and 0.72 dB, leading to an average coupling loss of 0.70 dB with a relatively small standard deviation of only 0.06 dB. The small variation of the losses is due to the fact that the active area of the Rx PD (32 µm diameter) is much larger than the spot of the received beam, which is focused down to a diameter of 12 µm on the PD facet by the associated FaML.

We also measure the lateral misalignment excess loss for one of the four nominally identical FaML-based fiber-PD connections of the Rx module, see Fig. 3(c). For simplicity, we only perform one-dimensional movements of the MT ferrule along the $x$-direction (left panel) and along the $+y$-direction (right panel). Note that the movement in $-y$-direction was restricted to avoid collision of the MT ferrule and the PD chip, see Fig. 2(c). For a 1 dB loss, a deviation of $\Delta x = \pm 62$ µm in $x$-direction and of $\Delta y = _{(}\pm_{)} 54$ µm in $y$-direction can be tolerated, where the boundary for a movement to the $-y$-direction was estimated from its counterpart in $+y$-direction. The position obtained by the machine-vision-based passive assembly procedure is indicated at $(x = 4 \text{µm}, y = 4 \text{µm})$, see grey tick label "passive". Again, the automated passive alignment loss is well inside the 1 dB tolerance, and no penalty could be quantified within the measurement accuracy. These measured coupling losses and alignment tolerances are slightly worse than the 0.5 dB that have previously been demonstrated both for actively and for passively aligned interfaces between PD and MM-FA [8,12], while the alignment tolerances are comparable.

We finally compare the performance of our approach to commercially available coupling schemes based on high-precision injection-molded plastic parts [10,25]. The most prominent example is the PRIZM® LightTurn® [10,24], which has become a widely used solution for coupling between fiber and device arrays with standard pitches of 250 µm. In terms of coupling losses, the 0.5 dB (Tx) and the 0.7 dB (Rx) demonstrated for our FaML-based passive assembly approach can well compete with the specified sub-2 dB losses of the PRIZM® LightTurn® [25]. The presumably overestimated standard deviations of 0.25 dB (Tx) and 0.06 dB (Rx) for the FaML-based approach also compares favorably with the standard deviation of 0.22 dB (Tx) and 0.34 dB (Rx) achieved with the PRIZM® LightTurn® [10]. In this context, it should be noted that the PRIZM® LightTurn® also requires precise alignment to ensure that the underlying socket is mounted to the PCB in the correct position with respect to the VCSEL / PD arrays. The alignment precision required in this step is related to the diameter of the beam at the corresponding VCSEL / PD facet, which is smaller than the beam diameter in the collimated free-space beam section between the FaML, dictating the alignment accuracy in the FaML approach. We hence believe that our approach can offer higher alignment tolerances along with comparable or lower losses, while offering a path towards parallel coupling to the individual cores of MCF, see next section.

Based on our experiments, indicating comparatively large alignment tolerances, we believe that coupling of larger fiber arrays to corresponding VCSEL or PD arrays should not represent a problem. These expectations are also supported by similar experiments that were performed with arrays of SMF and single-mode waveguides [29,44], which are much more sensitive to alignment errors. In these experiments [44], arrays of silicon photonic (SiP) waveguides with rather small mode-field diameters of the order of 2.5 µm were coupled to arrays of SMF, equipped with FaML consisting of TIR mirrors and focusing lens surfaces. It was found that the resulting position variations of the generated beam foci were dominated by the position variations of the SMF cores within the underlying fiber array. Variations of the 3D-printed FaML did not play a significant role, indicating the precise alignment of the FaML to the respective fiber cores during the fabrication process.

## 4. Coupling of VCSEL / PD to multiple cores of an MCF

In a second set of experiments, we demonstrate the viability of our concept by connecting linear VCSEL and PD arrays to MM-MCF. The associated coupling scheme of Fig. 1(c) is displayed in more detail for the Tx in Fig. 4(a) and for the Rx in Fig. 4(b). In Figure 4(a), three Tx VCSEL are coupled to three MM-MCF cores of the Tx fiber labelled ❶, ❷, and ❸, that lie in a common $z$-normal plane. Figure 4(b) shows the corresponding coupling of three Rx PD to three cores of the Rx fiber, labelled ❹, ❺, and ❻. The outer FaML ❶ and ❸ on the VCSEL as well as the outer FaML ❹ and ❻ on the PD are tilted. Figure 4(c) and (d) display technical drawings of the Tx coupling scheme ❷ and of the Rx coupling arrangement ❺, respectively, in an $x$-normal plane. For the Tx, Fig. 4(c), the refracting Surface S1 collimates the VCSEL beam to a MFD of 28 µm, measured half-way between the two FaML. The entrance Surface S2 of the center FaML on the Tx MM-MCF is located at a distance $d = 525 \mu m$ from the apex of the corresponding FaML on the VCSEL. Surface S2 is designed to reduce the MFD at the position of the TIR mirror S3 to 22 µm to avoid clipping due to the limited lateral size of the TIR mirror as dictated by the 39 µm pitch of the fiber cores. Refracting Surface S4 forms the beam such that the MM-MCF core is illuminated up to 70 % of the core radius $a = 13 \mu m$ and up to 70 % of the numerical aperture of 0.21. This excitation approximates an LPS launch [39] such that, in combination with a cladding-mode stripper, an under-estimation of the coupling loss is avoided. For the Rx, Fig. 4(d), a TIR mirror S5 redirects the beam towards the PD. The exit Surface S6 of the FaML at the MCF collimates the beam to a MFD of 28 µm, measured again half-way between the two FaML, while the input Surface S7 of the FaML at the PD focuses the expanded beam to a spot with a MFD of 15 µm, significantly smaller than the diameter of 32 µm of the light-sensitive PD area. The apices of Surfaces S6 and S7 are again separated by $d = 525 \mu m$. Figure 4(e) shows a technical drawing of the FaML ❸ and ❻ in a $y$-normal plane. The VCSEL / PD beam is redirected by an angled Surface S0, designed for a beam tilt angle $\alpha = 14.6°$ with respect to the $z$-axis.

A photograph of the transceiver module following the concept in Fig. 1(c) is depicted in Fig. 5(a). Figure 5(b) shows a close-up of three VCSEL / PD, each connected to three cores of a Tx / Rx MM-MCF, respectively. Inset (i) of Fig. 5(b) gives a magnified view of the FaML on the fiber facet containing TIR mirrors and beam-expander lenses. Inset (ii) of Fig. 5(b) depicts the FaML attached to the VCSEL array.

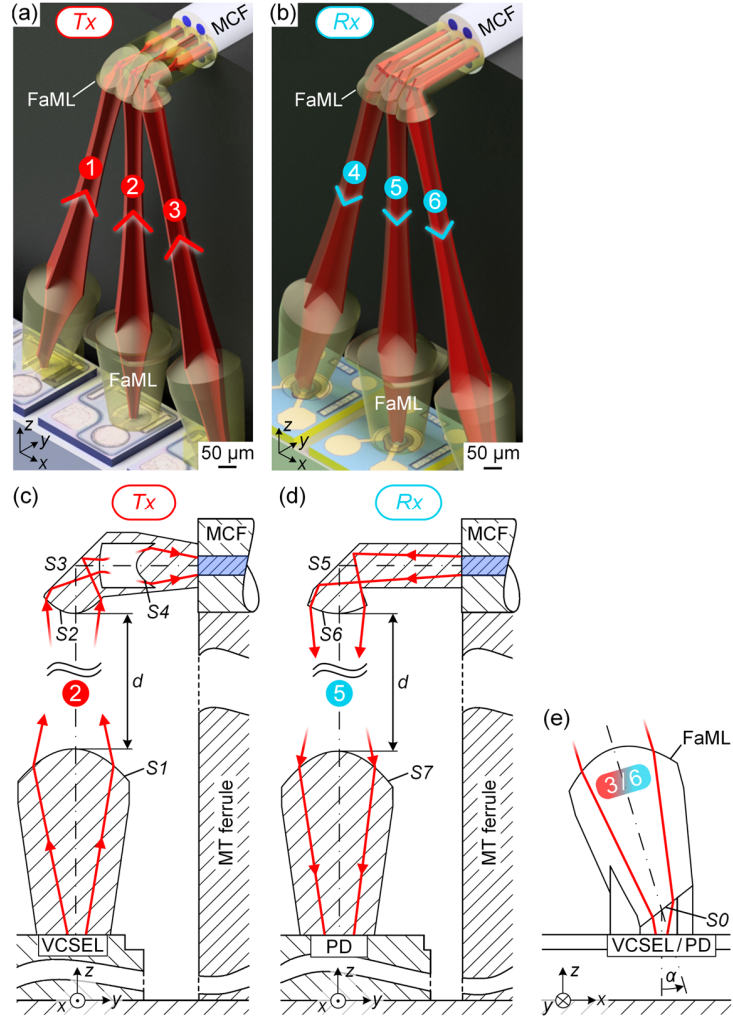

**Fig. 4.** Concept for coupling of VCSEL and PD arrays to MM-MCF as shown in Fig. 1(c) along with corresponding technical drawings. **(a)** Three VCSEL coupled to three MCF cores, labelled ❶, ❷, and ❸, that lie in a common $z$-normal plane. **(b)** Three PD coupled to three cores of the Rx fiber, labelled ❹, ❺, and ❻. **(c)** Technical drawing of the Tx coupling scheme ❷. The refracting Surface S1 collimates the VCSEL beam. The entrance refracting Surface S2 of the FaML on the MCF is located at a distance $d = 525\,\mu\mathrm{m}$. It reduces the MFD at the position of the TIR mirror S3 to avoid clipping that would occur due to the limited lateral size of the TIR mirror as dictated by the $39\,\mu\mathrm{m}$ pitch of the fiber cores. Refracting Surface S4 forms the beam for illuminating the MM-MCF core up to 70 % of the core radius, and up to 70 % of the numerical aperture for approximating an LPS launch [39]. **(d)** Technical drawing of the Rx coupling scheme ❺. A TIR mirror S5 redirects the beam towards the PD. The exit Surface S6 of the FaML at the MCF collimates the beam, and Surface S7 at the entrance face of the FaML at the PD focuses the beam to a spot, significantly smaller than the diameter of the light-sensitive PD area. The apices of Surfaces S6 and S7 are again separated by $d = 525\,\mu\mathrm{m}$. **(e)** Technical drawing of FaML ❸ and ❻ viewed in $y$-direction. The outer Tx FaML ❶, ❸ and the Rx FaML ❹, ❻ are tilted. The VCSEL / PD beam is directed by an angled Surface S0, designed for a beam tilt angle $\alpha = 14.6°$ with respect to the $z$-axis.

### 4.1 Module assembly

In contrast to the assemblies discussed in Section 3, where Tx and Rx fibers were mounted into separate MT ferrules, the MM-MCF scheme relies on a single MT ferrule that contains both the Tx and the Rx MCF, see Fig. 5(a). As a consequence, the VCSEL / PD arrays need to be mounted to the PCB collinearly with the correct pitch. Specifically, the distance between the center VCSEL and PD elements ❷ and ❺ needs to match the distance of 1750 µm between the central cores of the Tx and the Rx MM-MCF, see Fig. 5(b). Given the rather high positioning tolerances of the MM coupling interfaces, this step is well manageable using state-of-the-art pick-and-place equipment. Apart from this, the assembly process for the MM-MCF scheme relies on the same machine and is largely similar to the procedure used for the Tx and the Rx modules presented in Section 3.1. In a first step, we grip the MT ferrule and perform a tilt correction to align the plane defined by the fiber axes exactly parallel to the plane in which the PD and VCSEL arrays are mounted. We then use camera-based machine vision to detect the positions of the FaML at the Tx VCSEL ❷ and Rx PD ❺, exploiting a collar that surrounds the FaML to improve the accuracy of the image recognition, see Fig. 5(b). We then extract the connecting line between VCSEL ❷ and PD ❺ as a reference for aligning the MT ferrule. The MT ferrule is then moved in $x$- and $y$-direction for adjusting the line M1−M2 between the 3D-printed marker holes M1, M2 vertically above the connecting line between VCSEL ❷ and PD ❺, Fig. 5(b). For lateral alignment, we extract the mid-point of the connecting line between markers M1 and M3 to locate the position of the apex belonging to FaML ❷ on the fiber side. This apex point is then laterally shifted along the M1−M3 connection to position it exactly vertically above the FaML on top of VCSEL ❷. The $z$-position of the MT ferrule is then adjusted to provide the designed distance of $d = 525\,\mu m$ between the apices of the center FaML on VCSEL ❷ and PD ❺ and the corresponding apex on the MCF FaML, see Fig. 4(c) and (d). A low-shrinkage UV-curable epoxy glue (EMI Optocast 3410 Gen2) is used at the four corners of the MT ferrule to fix its final position.

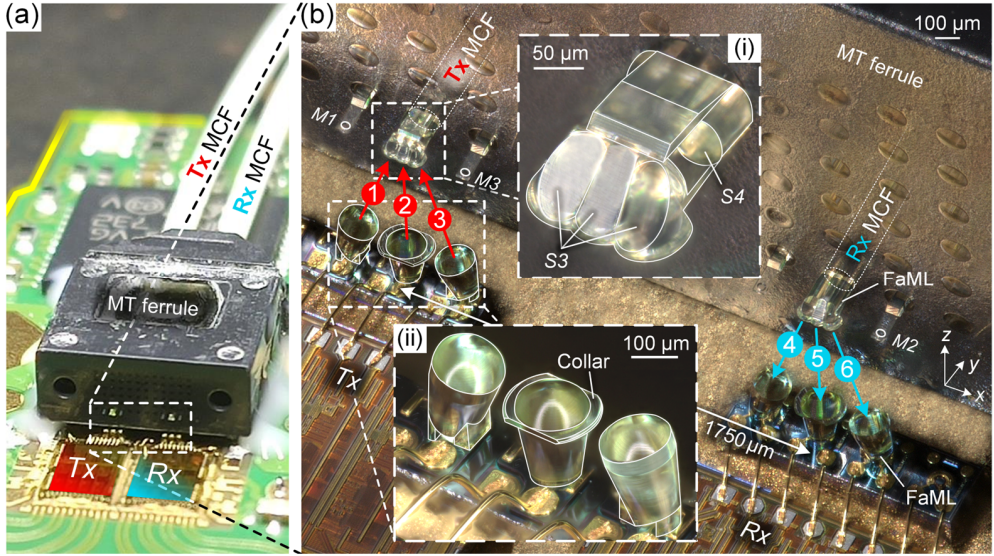

**Fig. 5.** Three-channel transceiver module implemented on a small form-factor pluggable PCB. Three Tx VCSEL ❶, ❷, and ❸ and three Rx PD ❹, ❺, and ❻ are connected to three cores of the associated MM-MCF. **(a)** Overview photograph corresponding to schemes illustrated in Fig. 4(a) and (b). **(b)** Micrograph of the optical couplers. Inset (i) gives a magnified view of the fiber-attached optical elements, see technical drawing in Fig. 4(c) for details. Inset (ii) depicts the expander lenses attached to the three VCSEL, and the "collar" supporting machine vision.

*4.2 Alignment tolerance and coupling loss*

For quantifying the alignment tolerances, we repeat the experiments described in Section 3.2, where we move the MT ferrule in $x$- and $y$-direction prior to applying the glue. Note that due to the expanded spot size of the beams a precise alignment along the beam direction ($z$-direction) is again not required. By recording the power in the three cores ❶, ❷, and ❸ of the Tx MCF and by measuring the photocurrents of the three Rx PD ❹, ❺, and ❻, we extract the respective excess loss at the Tx and Rx coupling interfaces for lateral displacements along the *x*- and +*y*-direction, see Fig. 6(a) and (c). The average optimum position maximizing the sum of the powers in all three Tx channels is at $(x = 0\,\mu\text{m}, y = 0\,\mu\text{m})$, and the excess losses of the individual Tx and Rx channels are indicated in relation to the loss of the respective channel found at this position. For a 1 dB excess loss for the Tx coupling, a deviation of $\Delta x = \pm 18\,\mu\text{m}$ in *x*-direction and of $\Delta y = [\pm]13\,\mu\text{m}$ in *y*-direction can be tolerated, where the movement to negative *y*-coordinates was again restricted to avoid collision of the MT ferrule and the VCSEL chip, see Fig. 4(a), and where the boundary for a movement to the –*y*-direction was estimated from its counterpart in +*y*-direction. For passive alignment of the MT ferrule, we find offsets of $x = 1\,\mu\text{m}$ measured along the longer side of the VCSEL chip base and of $y = 4\,\mu\text{m}$ measured along the shorter side – these positions are again marked by the tick labels "passive" in Fig. 6(a) and (c). The offset along the *y*-direction is again larger than the offset along the *x*-direction, which we attribute to the fact that the tilt measurement of the VCSEL chip about its long axis is subject to higher uncertainties, see Section 3.2 for a more detailed discussion.

Using active alignment, the average power in the Tx MCF cores ❶, ❷, and ❸ reaches 1.93 dBm for a bias current of 3.3 mA applied to the VCSEL. This corresponds to an absolute average coupling loss of 0.67 dB. We estimate absolute coupling losses of 1.38 dB, −0.16 dB, and 0.93 dB for VCSEL ❶, ❷, and ❸ as defined in Fig. 5(b), using again the average VCSEL emission power as a reference. This leads to the reported average coupling loss of 0.67 dB with a standard deviation of 0.79 dB. We attribute the again rather high standard deviation as well as the unphysical negative dB-value of the coupling efficiency for VCSEL ❷ to the fact that the reported coupling efficiencies are subject to the unknown variations of the underlying VCSEL emission powers, which could not be measured individually, see Section 3.2 for a more detailed explanation. Note also that the outer connections ❶ and ❸ require one additional lens Surface S0 for redirection of the beams, see Fig. 4(e), which is a possible reason for the increased loss compared to the central connection ❷.

For the automated passive alignment of our assembly, we measure an average power of 1.59 dBm, which corresponds to an average coupling loss of 1.0 dB. The absolute coupling losses for the individual VCSEL-MCF connections amount to 1.55 dB, 0.16 dB, and 1.46 dB for VCSEL ❶, ❷, and ❸, respectively, taking again the average VCSEL emission power as a reference. The passive alignment penalties for the individual channels hence amount to 0.17 dB for VCSEL ❶, to 0.32 dB for VCSEL ❷, and to 0.53 dB for VCSEL ❸, leading to an average penalty of 0.33 dB.

For measuring the absolute losses of the Rx PD, we use an MCF connector to inject a known power into the three relevant cores of the Rx MCF and measure the photocurrents of the respective Rx PD. The photocurrents are translated into optical power levels using the data-sheet specification of the responsivity *S*. For a 1 dB excess loss, a deviation of $\Delta x = \pm 23\,\mu\text{m}$ in *x*-direction and of $\Delta y = [\pm]25\,\mu\text{m}$ in *y*-direction can be tolerated, where the movement to negative *y*-coordinates was again restricted to avoid collision of the MT ferrule and the PD chip, see Fig. 4(b), and where the boundary for a movement to the –*y*-direction was estimated from its counterpart in +*y*-direction. The position-dependent excess losses for the three channels ❹, ❺, and ❻ exhibit plateaus due to the fact that the 15 µm spot size generated by the FaML on the PD surface is smaller than the 32 µm diameter of the active PD area. The excess losses of the three Rx channels shown in Fig. 6(c) are again measured with respect to the absolute loss of the respective channel found at the optimum coupling position $(x = 0\,\mu\text{m}, y = 0\,\mu\text{m})$. The coupling losses of the individual MCF-PD connections at this position amount to 1.32 dB for PD ❹, to 0.40 dB for PD ❺, and to 0.30 dB for PD ❻, leading to an average coupling loss of 0.63 dB with a standard deviation of 0.56 dB. For the automated passive alignment of our assembly, no penalty could be quantified within the measurement accuracy. This leads to equal coupling losses for the optimum Tx position $(x = 0\,\mu\text{m}, y = 0\,\mu\text{m})$ and for the passively aligned position, indicated again by tick marks "passive", see Fig. 6(c).

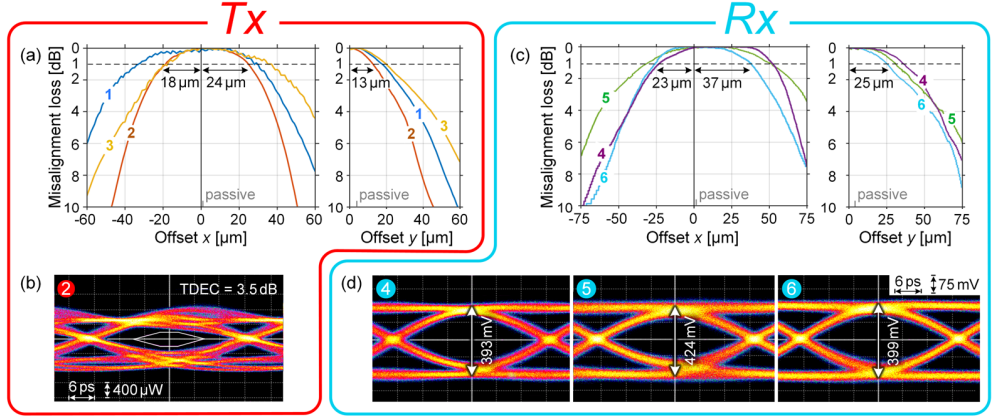

**Fig. 6.** Alignment tolerances and eye diagrams of a three-lane transmitter (Tx, VCSEL) ❶, ❷, ❸, and a three-lane receiver (Rx, PD) ❹, ❺, ❻, each lane coupled to one of three cores of a MCF, Fig. 4. We extract the respective excess loss at the Tx and Rx coupling interfaces for lateral displacements along the *x*- and +*y*-direction. The average optimum position maximizing the sum of the powers in all three Tx channels is at $(x = 0\,\mu m, y = 0\,\mu m)$, and the excess losses of the individual Tx and Rx channels are indicated in relation to the loss of the respective channel found at this position. The excess loss for an automated passive assembly is indicated by the tick labels "passive". **(a)** Lateral misalignment excess loss measured at the Tx for the three FaML-based VCSEL-MCF connections. For simplicity, we restrict our experiment to linear movements along the *x*-direction (left panel) and along the +*y*-direction (right panel). For a 1 dB excess loss, a deviation of $\Delta x = \pm 18\,\mu m$ in *x*-direction and of $\Delta y = [\pm]13\,\mu m$ in *y*-direction can be tolerated. Note that the movement to negative *y*-coordinates was restricted to avoid collision of the MT ferrule and the VCSEL / PD chip, Fig. 4 (c) and (d), and that the boundary for a movement to the –*y*-direction was therefore estimated from its counterpart in +*y*-direction. The average excess loss for passive automated assembly at $(x = 1\,\mu m, y = 4\,\mu m)$ is 0.33 dB. At the optimum position, we measure an absolute average coupling loss of 0.67 dB for the three Tx MCF cores ❶, ❷, ❸. For the automated passive alignment, the absolute average coupling loss amounts to 1.0 dB. **(b)** Characterization of the Tx with on-off-keying (OOK) signals (PRBS31) at line rates of 25.78125 Gbit/s, which are coupled to the Tx driver IC, see Fig. 5. Each VCSEL is operated at a bias current of 3.3 mA and a peak-to-peak modulation current of 4 mA. The optical output from cores ❶, ❷, and ❸ of the Tx fiber is sent to a sampling oscilloscope. As an example, we show the eye diagram measured from Tx core ❷ while the two other Tx VCSEL ❶ and ❸ are in operation. The transmitter-and-dispersion eye-closure (TDEC) penalty is 3.5 dB, and an extinction ratio of 3.7 dB is measured. **(c)** Lateral misalignment excess loss measured at the Rx for the three FaML-based MCF-PD connections. For simplicity, we restrict our experiment to linear movements along the *x*-direction (left panel) and along the +*y*-direction (right panel). For a 1 dB excess loss, a deviation of $\Delta x = \pm 23\,\mu m$ in *x*-direction and of $\Delta y = [\pm]25\,\mu m$ in *y*-direction can be tolerated, where the boundary for a movement to the –*y*-direction was again estimated from its counterpart in +*y*-direction. The position-dependent excess losses for the three channels ❹, ❺, and ❻ exhibit plateaus due to the fact that the 15 µm spot size generated by the FaML on the PD surface is smaller than the 32 µm diameter of the active PD area. As a consequence, the measured absolute average loss of the three Rx channels amounts to 0.63 dB both for the optimum Tx position at $(x = 0\,\mu m, y = 0\,\mu m)$ and for the passively aligned position, indicated again by tick marks "passive". **(d)** Characterization of the Rx with OOK signals at a line rate of 25.78125 Gbit/s. The signal from an optical transmitter with known specifications is fed to the three cores ❹, ❺, and ❻ of the Rx MCF, detected by the Rx PD, and the output signals of the Rx driver IC are recorded by a sampling oscilloscope. The eye diagrams for Rx ❹, ❺, and ❻ are shown, indicating the peak-to-peak voltage of the corresponding receiver.

To the best of our knowledge, these experiments represent the first demonstration of a coupling interface that connects individual cores of an MCF to VCSEL / PD arranged in a standard linear array without the need for additional fiber-based or waveguide-based fan-out structures. Other approaches relying on custom device arrays in 2D arrangements matched to the cross-section of the respective MCF have been pursued [17−21]. Such demonstrations, however, still rely on active alignment, leading, e.g., to minimum coupling losses of 0.98 dB between VCSEL and a seven-core MM-MCF [20]. These losses are slightly higher than the 0.67 dB and the 0.63 dB achieved by active alignment in our experiments for the Tx and Rx, respectively, and our passively assembled module can still well compare to these losses – besides avoiding non-standard device arrangements that are adapted to the cross section of the respective MCF. It should be noted that coupling schemes based on custom device arrays require a generally denser spacing of the VCSEL / PD compared to the traditional 250 µm pitch, which may limit the high-speed performance of the devices due to higher temperatures [18].

Using FaML-based free-space fan-out schemes to address the individual cores of a MCF can also open a path towards connecting two-dimensional VCSEL or PD arrays to one-dimensional arrays of MCF. We believe that this approach has the potential to greatly increase the number of parallel channels per transceiver, while maintaining the fiber installations manageable. For the present work, such a demonstration was not possible due to the lack of 2D arrays of VCSEL and PD and of associated read-out electronics that can individually address the various devices in such a matrix.

*4.3 Data-transmission experiments*

To demonstrate the viability of the presented concept, the assembled $3 \times 25$ Gbit/s transceiver module was characterized by measuring the transmitter-and-dispersion eye-closure (TDEC) penalty, the Tx power per lane, the optical modulation amplitude (OMA), the extinction ratio, and the Rx power per lane – similar to the procedures described in the IEEE 802.3 industry standard [45]. To this end, we use the SFP interfaces on the PCB to feed the Tx driver IC with three on-off-keying (OOK) signals at a line rate of 25.78125 Gbit/s. The drive signals were derived from a pseudorandom binary sequence of length $2^{31}-1$ (PRBS31). We set the bias current of the VCSEL to 3.3 mA and use a current modulation of $\pm 2$ mA, i.e., a peak-to-peak swing of 4 mA, for the OOK signals. The optical output from cores ❶, ❷, and ❸ of the Tx fiber, see Fig. 5, is sent to a sampling oscilloscope. Figure 6(b) shows an exemplary eye diagram measured from Tx core ❷ with the two other Tx VCSEL ❶ and ❸ in operation. We also evaluated the optical crosstalk by measuring the power coupled from Tx VCSEL ❶ to, e.g., Tx core ❷. All such experiments lead to negligible power readings in the "unwanted" Tx cores, indicating that the optical crosstalk was below our measurement sensitivity. From the recorded eye diagram, a TDEC of 3.5 dB is found, which is clearly below the maximum value of 4.3 dB specified for 100GBASE-SR4 transceivers in the IEEE 802.3 standard [45]. The average Tx power amounts to 1.56 mW (1.9 dBm) while an OMA of 1.35 mW (1.1 dBm) is measured – both of these values are well within the respective range specified in the IEEE 802.3 standard [45]. From these numbers, we extract a ratio of the OMA to the TDEC of 0.58 mW or –2.4 dBm, which is well above the minimum value of –7.3 dBm specified for the "launch power in OMA minus TDEC"-parameter in the IEEE 802.3 standard [45]. Finally, an extinction ratio of 3.7 dB and an average off-state transmitter power of –36 dBm are measured, which also fulfill the requirements imposed by the IEEE 802.3 standard [45].

For the receiver characterization, the signal from an optical transmitter with known specifications is fed to the three cores ❹, ❺, and ❻ of the Rx MCF. The signal is then detected by the PD, and the output signals of the Rx driver IC are recorded by a sampling oscilloscope. Figure 6(d) shows the eye diagrams for Rx ❹, ❺, and ❻, indicating the peak-to-peak voltage of the corresponding receiver. The average received power per lane amounts to 1.07 mW, 1.39 mW, and 1.18 mW, again fulfilling the respective specification of the IEEE 802.3 standard [45].

## 5. Summary

We demonstrate that 3D-printed facet-attached microlenses (FaML) open an attractive path for connecting multimode fiber arrays (MM-FA) as well as individual cores of multimode multicore fibers (MM-MCF) to standard arrays of vertical-cavity surface-emitting lasers (VCSEL) or photodiodes (PD) with pitches of 250 µm. The FaML, which can be printed by high-precision multi-photon lithography directly on the device and fiber facets, are designed to collimate the associated beams to large diameters of tens of micrometers, thereby greatly relaxing alignment tolerances in both the transverse and axial direction. To demonstrate the viability of the proposed concepts, we further perform a series of proof-of-concept experiments using a custom pick-and-place machine to mount the FaML-equipped fiber arrays to the PCB in a fully automated process, controlled by machine vision and height measurements. Using active alignment, we show connections between VCSEL / PD arrays and MM-FA, achieving average coupling losses as low as 0.35 dB for the Tx and 0.70 dB for the Rx along with lateral 1 dB alignment tolerances of ± 17 µm (Tx) and ± 62 µm (Rx), respectively. When using the machine-vision based passive alignment process, we demonstrate average coupling losses of 0.50 dB (Tx) and 0.70 dB (Rx). To the best of our knowledge, these results are among the lowest losses and the highest alignment tolerances so far demonstrated for coupling between VCSEL / PD arrays and MM-FA. We further connect a linear VCSEL / PD array to distinct cores of a single MCF. When using active alignment, we achieve average coupling losses of 0.67 dB (Tx) and 0.63 dB (Rx) along with lateral 1 dB alignment tolerances of ± 18 µm (Tx) and ± 25 µm (Rx), respectively. Using the machine-vision based passive alignment process, we achieve average coupling losses of 1.0 dB (Tx) and 0.63 dB (Rx). To the best of our knowledge, these experiments represent the first demonstration of a coupling interface that connects individual cores of an MCF to VCSEL / PD arranged in a standard linear array without the need for additional fiber-based or waveguide-based fan-out structures. Using this approach, we finally built a 3 × 25 Gbit/s transceiver assembly which fits into a small form-factor pluggable module and which fulfills many performance metrics specified in the IEEE 802.3 standard. We believe that 3D-printed FaML could pave a path towards highly scalable transceiver assemblies that exploit readily available VCSEL and PD arrays in combination with parallel transmission through multimode multicore fibers without the need for expensive multi-step assembly procedures or technically complex fan-out structures.

**Appendix**

*A. Beam quality measurements*

For designing the facet-attached microlenses (FaML), multimode light propagation is emulated by using Gaussian beams with an effective wavelength $\lambda_{\text{eff}} = M^2 \lambda$ that is increased with respect to the true vacuum wavelength $\lambda$ of the underlying device by the measured beam quality factor $M^2$. In the following, we shortly sketch the mathematical background of this approach, the foundations of which are explained in more detail in [46−55].

One of the most general mathematical descriptions valid for all types of coherent and non-coherent optical beams relies on the so-called Wigner distribution function (WDF), first introduced by Wigner in 1932 in the context of quantum states represented as a distribution in terms of both position and momentum [53]. In the following, we assume Cartesian coordinates ($x, y, z$), where the beam axis corresponds to the $z$-axis without loss of generality, and where the transverse position vector is given by $\vec{r}_t = (x, y)^T$. For the most general case, the complex-valued components $\vec{E}(x, y, z)$ of the electric field are non-stationary stochastic processes in $x$ and $y$, and the associated coherence function (autocorrelation) $\Gamma(\vec{r}_{t,1}, \vec{r}_{t,2}; z)$ depends individually on both transverse positions $\vec{r}_{t,1}$ and $\vec{r}_{t,2}$ rather than on the difference $\vec{r}_{t,1} - \vec{r}_{t,2}$ only. The WDF is obtained by re-writing the coherence function as $\Gamma(\vec{r}_t + \vec{r}_t\,'/2, \vec{r}_t - \vec{r}_t\,'/2; z)$ with position vector $\vec{r}_t = (x, y)^T$ and offset vector $\vec{r}_t\,' = (x', y')^T$ and by computing the two-dimensional Fourier transform with respect to $\vec{r}_t\,'$,

$$\tilde{W}(\vec{r}_t, \vec{k}_t; z) = \iint \Gamma\left(\vec{r}_t + \frac{\vec{r}_t\,'}{2}, \vec{r}_t - \frac{\vec{r}_t\,'}{2}; z\right) e^{-j\vec{k}_t^T \cdot \vec{r}_t\,'} \, dx'dy'. \quad (1)$$

The corresponding components $k_{t,x}$ and $k_{t,y}$ of the transverse spatial-frequency vector $\vec{k}_t$ can be interpreted as transversal projection $\vec{k}_t = (k_{t,x}, k_{t,y})^T = k\vec{\Theta}_t$ of a paraxial wave-propagation vector $\vec{k}$, where $k = \omega n/c = 2\pi n/\lambda$ denotes the wavenumber in the respective medium with refractive index $n$ and where $\vec{\Theta}_t = (\theta_x, \theta_y)^T$ indicates the direction of the transverse wave-vector components. One can then rewrite Eq. (1) as

$$W\left(\begin{pmatrix}x\\y\end{pmatrix},\begin{pmatrix}\theta_x\\\theta_y\end{pmatrix};z\right) = \iint \Gamma\left(\begin{pmatrix}x\\y\end{pmatrix}+\frac{1}{2}\begin{pmatrix}x'\\y'\end{pmatrix},\begin{pmatrix}x\\y\end{pmatrix}-\frac{1}{2}\begin{pmatrix}x'\\y'\end{pmatrix};z\right) e^{-jk(x'\theta_x+y'\theta_y)}\,\mathrm{d}x'\mathrm{d}y', \qquad (2)$$

where we use the notation $W$ without the tilde to denote the WDF in terms of the arguments $\theta_x$ and $\theta_y$. While it is difficult to assign a physical meaning to the WDF itself, the corresponding marginal distribution obtained by integration over $\theta_x$ and $\theta_y$ can be interpreted as the spatial intensity distribution $I(x,y;z)$,

$$I(x,y;z) = \left(\frac{k}{2\pi}\right)^2 \iint W\left(\begin{pmatrix}x\\y\end{pmatrix},\begin{pmatrix}\theta_x\\\theta_y\end{pmatrix};z\right)\,\mathrm{d}\theta_x\,\mathrm{d}\theta_y. \qquad (3)$$

The total power $P$ of the beam is obtained by integrating the spatial intensity distribution over the two transverse coordinates $x$ and $y$,

$$P(z) = \iint I(x,y;z)\,\mathrm{d}x\mathrm{d}y = \left(\frac{k}{2\pi}\right)^2 \iiiint W\left(\begin{pmatrix}x\\y\end{pmatrix},\begin{pmatrix}\theta_x\\\theta_y\end{pmatrix};z\right)\,\mathrm{d}\theta_x\,\mathrm{d}\theta_y\,\mathrm{d}x\mathrm{d}y. \qquad (4)$$

In analogy to Eqs. (3) and (4), the marginal distribution obtained by integration over $x$ and $y$ corresponds to the $z$-dependent spatial power spectrum, and the total power of the beam is again obtained by additionally integrating over $\theta_x$ and $\theta_y$. With the previous definitions in Eqs. (2)–(4), one can then define a normalized average value $\bar{f}(z)$ of any function $f(x,y,\theta_x,\theta_y)$ weighted by the WDF,

$$\bar{f}(z) = \frac{1}{P}\left(\frac{k}{2\pi}\right)^2 \iiiint f(x,y,\theta_x,\theta_y) W\left(\begin{pmatrix}x\\y\end{pmatrix},\begin{pmatrix}\theta_x\\\theta_y\end{pmatrix};z\right)\,\mathrm{d}\theta_x\,\mathrm{d}\theta_y\,\mathrm{d}x\mathrm{d}y. \qquad (5)$$

In the following, we assume that the propagation of light in our assemblies can be described by so-called aligned simple astigmatic (ASA) beams, which is a common assumption for laser beams [54]. For ASA beams, the $x$- and the $y$-axis of the coordinate system can be chosen such that the WDF $W((x,y)^\mathrm{T},(\theta_x,\theta_y)^\mathrm{T};z)$ can be separated and expressed by a product of two functions $f_1(x,\theta_x;z)$ and $f_2(y,\theta_y;z)$, each of which depends only on one of the transverse directions $x$ or $y$ and on the corresponding transverse components $\theta_x$ or $\theta_y$ of the direction vector $\vec{\Theta}_\mathrm{t}$. In the case of light emitted by a laser, an ASA beam can be thought of as a beam with elliptical intensity distributions in the transverse plane, where the major and minor axis of the ellipses are aligned along the $x$ and the $y$-direction. The second central moments $\sigma_x^2(z)$ and $\sigma_y^2(z)$ of the intensity distribution $I(x,y;z)$ can then be found by adopting Eq. (5) accordingly,

$$\sigma_x^2(z) = \frac{\iint (x-\bar{x})^2 I(x,y;z)\,\mathrm{d}x\mathrm{d}y}{\iint I(x,y;z)\,\mathrm{d}x\mathrm{d}y}, \quad \sigma_y^2(z) = \frac{\iint (y-\bar{y})^2 I(x,y;z)\,\mathrm{d}x\mathrm{d}y}{\iint I(x,y;z)\,\mathrm{d}x\mathrm{d}y}, \qquad (6)$$

where

$$\bar{x} = \frac{\iint x\,I(x,y;z)\,\mathrm{d}x\mathrm{d}y}{\iint I(x,y;z)\,\mathrm{d}x\mathrm{d}y}, \quad \bar{y} = \frac{\iint y\,I(x,y;z)\,\mathrm{d}x\mathrm{d}y}{\iint I(x,y;z)\,\mathrm{d}x\mathrm{d}y}, \qquad (7)$$

denote the first moments of the intensity distribution $I(x,y;z)$. Assuming beams that propagate along the $z$-direction within the limitations of the paraxial approximation, it can be shown [52] that the second central moments $\sigma_x^2(z)$ and $\sigma_y^2(z)$ of the intensity distributions as given by Eq. (6) evolve according to a simple quadratic relationship,

$$\sigma_x^2(z) = \sigma_{0,x}^2 + \sigma_{\theta_x}^2 (z - z_{0,x})^2, \quad \sigma_y^2(z) = \sigma_{0,y}^2 + \sigma_{\theta_y}^2 (z - z_{0,y})^2. \tag{8}$$

The two expressions in Eq. (8) are governed by an overall six parameters: The waist positions $z_{0,x}$ and $z_{0,y}$, the associated variances $\sigma_{0,x}^2$ and $\sigma_{0,y}^2$ of the WDF or, equivalently, of the intensity distribution along the $x$ and the $y$-direction, and two additional parameters $\sigma_{\theta_x}^2$ and $\sigma_{\theta_y}^2$ that describe the divergence of the beam in $x$ and $y$-direction, respectively.

The previous relations can also be applied to a fundamental Gaussian $TEM_{00}$ beam, i.e., a monochromatic Gaussian beam containing only the fundamental Hermite–Gaussian mode, propagating under the restrictions of paraxial optics. In this case, the divergences in the $x$- and $y$-direction are directly linked to the corresponding variances $\sigma_{G0,x}^2$ and $\sigma_{G0,y}^2$ of the intensity distributions in the respective waist at $z = z_{0,x}$ and $z = z_{0,y}$,

$$\sigma_{G,x}^2(z) = \sigma_{G0,x}^2 + \left(\frac{\lambda}{4\pi n \sigma_{G0,x}}\right)^2 (z - z_{0,x})^2,$$

$$\sigma_{G,y}^2(z) = \sigma_{G0,y}^2 + \left(\frac{\lambda}{4\pi n \sigma_{G0,y}}\right)^2 (z - z_{0,y})^2. \tag{9}$$

This leads to a constant product of the standard deviations $\sigma_{G0,x}, \sigma_{G0,y}$ and of corresponding divergence parameters $\tan \psi_{G,x}, \tan \psi_{G,y}$,

$$\sigma_{G0,x} \tan \psi_{G,x} = \sigma_{G0,y} \tan \psi_{G,y} = \frac{\lambda}{4\pi n}, \tag{10}$$

where

$$\tan \psi_{G,x} = \lim_{z \to \infty} \frac{\sigma_{G,x}(z)}{z} = \frac{\lambda}{4\pi n \sigma_{G0,x}}, \quad \tan \psi_{G,y} = \lim_{z \to \infty} \frac{\sigma_{G,y}(z)}{z} = \frac{\lambda}{4\pi n \sigma_{G0,y}}. \tag{11}$$

Note that the variances $\sigma_{G,x}^2$ and $\sigma_{G,y}^2$ specified in Eq. (9) relate to the radius over which the Gaussian intensity profile has dropped by a factor of $1/e$ compared to the on-axis maximum of the intensity. These variances can be translated into more widely used definitions of beam radii $w_{G,x} = 2\sigma_{G,x}$ and $w_{G,y} = 2\sigma_{G,y}$, which refer to the $1/e^2$ intensity points. Similarly, the beam divergence parameters $\tan \psi_{G,x}$ and $\tan \psi_{G,y}$ refer to the half-angle of an elliptical cone defined by the $1/e$ intensity contour lines, whereas it is more common to specify beam divergences $\tan \theta_{G,x} = 2 \tan \psi_{G,x}$ and $\tan \theta_{G,y} = 2 \tan \psi_{G,y}$ that relate to the $1/e^2$ contours.

For multimode beams, the direct connection between the standard deviations and divergence parameters according to Eq. (10) does not apply anymore. Still, it can be shown that the product of the divergence parameters $\tan \psi_x, \tan \psi_y$ and the standard deviations $\sigma_{0,x}, \sigma_{0,y}$ of the intensity distribution in the respective beam waist is an invariant property of the beam which is conserved by any lossless transformation of the beam, e.g., through lenses or mirrors [55]. The behavior of the multimode beam can thus be approximated by using a virtual fundamental Gaussian $TEM_{00}$ beam with an effectively increased wavelength $\lambda_{eff} = M^2 \lambda$ that leads to the same product of divergence parameter and corresponding variance of the intensity distribution. The beam-quality factor $M^2 = \lambda_{eff}/\lambda \geq 1$ is then a measure of the multimodedness of the beam. This consideration can be done separately for the $x$ and the $y$-direction, which may result in separate beam quality factors $M_x^2$ and $M_y^2$,

$$\sigma_{0,x} \tan \psi_x = M_x^2 \frac{\lambda}{4\pi n}, \quad \sigma_{0,y} \tan \psi_y = M_y^2 \frac{\lambda}{4\pi n}. \tag{12}$$

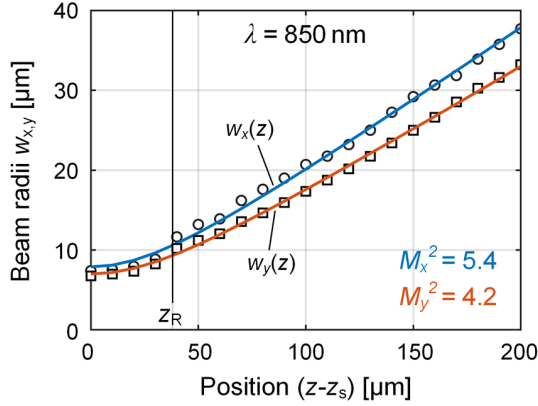

**Fig. 7.** Result of the VCSEL beam characterization. The measurements were taken using a 100×/0.8 microscope objective. The round and the square markers depict the beam radii $w_x = 2\sigma_x$ and $w_y = 2\sigma_y$ as a function of the axial position $(z-z_s)$ where $z_s$ corresponds to the surface of the VCSEL. Solid lines indicate a data fit to Eq. (8), resulting in beam quality factors $M_x^2 = 5.4$ and $M_y^2 = 4.2$ for beam waist radii $w_{0,x} = w_{0,y} = 7\,\mu m$ at $z_{0,x} = z_{0,y} = 0\,\mu m$. The average Rayleigh length amounts to $z_R = 38\,\mu m$ as indicated by the vertical line.

The only remaining task is now to extract the quality factors $M_x^2$ and $M_y^2$ of the beams emitted by our devices. To this end, we measure the intensity profiles $I(x, y; z)$ in a series of positions along the propagation direction $z$ using a microscope objective and a camera. We drive the VCSEL and move it with respect to the fixed objective using a motorized stage. All measurements are performed at a bias current of 3.3 mA, which was also used for the coupling-loss and data-transmission experiments, see, e.g., Section 3.2. From the recorded beam intensity profiles $I(x, y; z)$, we extract the second central moments $\sigma_x^2$ and $\sigma_y^2$ in $x$ and $y$-direction according to Eq. (6). In the measurement, sensor saturation must be avoided, and any background illumination needs to be subtracted. The waist positions $z_{0,x}$ and $z_{0,y}$, the associated variances $\sigma_{0,x}^2$ and $\sigma_{0,y}^2$ of the intensity distribution, and the beam-divergence parameters $\sigma_{\theta_x}^2$ and $\sigma_{\theta_y}^2$ are then extracted via a parameter fit using Eq. (8). The beam quality factors $M_x^2$ and $M_y^2$ are then given by $M_x^2 = (4\pi n/\lambda)\sigma_{0,x}\tan\psi_x$ and $M_y^2 = (4\pi n/\lambda)\sigma_{0,y}\tan\psi_y$, where the vacuum wavelength amounts to $\lambda = 850$ nm.

The result of the VCSEL beam characterization is shown in Fig. 7, where we plot the beam radii $w_x = 2\sigma_x$ (round markers) and $w_y = 2\sigma_y$ (square markers) as a function of the axial position $(z-z_s)$, where $z_s$ corresponds to the surface of the VCSEL. Fitting the data to Eq. (8) yields the plots shown in blue and red, characterized by $M_x^2 = 5.4$ and $M_y^2 = 4.2$ for beam-waist radii of $w_{0,x} = w_{0,y} = 7\,\mu m$ at $z_{0,x} = z_{0,y} = 0\,\mu m$. The average Rayleigh length for the profiles in $x$ and in $y$-direction amounts to $z_R = 38\,\mu m$ as indicated by the vertical line.

### B. Microlens simulations

For simulation of the FaML, we use an in-house developed simulation software based on the scalar wide-angle unidirectional wave-propagation method for step-index structures [36]. For all lens surfaces in our experiments, we use a rotational symmetric even-order polynomial with three free parameters $c_0, c_2$ and $c_4$ to represent the lens surface height above the $(x, y)$-plane,

$$h(r) = c_0 + c_2 r^2 + c_4 r^4 + ..., \qquad r = \sqrt{x^2 + y^2}. \tag{13}$$

Simulations are carried out by using the effective beam quality factor $M_{\text{eff}}^2 = \sqrt{M_x^2 \times M_y^2} = 4.8$, which is obtained as the geometrical mean [54] of the measured beam quality factors $M_x^2 = 5.4$ and $M_y^2 = 4.2$ along the respective principle axis of the beam, see Appendix A. Note that this simplified description by a single effective beam quality factor $M_{\text{eff}}^2$ in fact implies treating the beam as a stigmatic beam which is rotationally symmetric with respect to the $z$-axis [50].

This assumption is backed by the fact that the average ellipticity parameter $\bar{\varepsilon} = (1/N)\sum_{i=1}^{N} w_y(z_i)/w_x(z_i)$ over the $N = 21$ measured $z$-positions is larger than 0.87 and the beam profiles may therefore be considered to be of circular symmetry according to the ISO/IEC 11146-1 standard [54]. The multimode beam at vacuum wavelength $\lambda = 850\,\text{nm}$ is consequently emulated by a single effective wavelength $\lambda_{\text{eff}} = \lambda M_{\text{eff}}^2 = 4.08\,\mu\text{m}$ [49,51].

*C. Fabrication*

The FaML in our assemblies discussed in Section 3 and 4 are printed separately to the VCSEL / PD chips and the facet of the MT ferrule using high-resolution multi-photon lithography [30]. For convenience, the two printing steps have been carried out simultaneously on separate machines with a known-good set of parameters for the respective photoresist used.

The FaML on the MT ferrule were fabricated using an in-house-built lithography system with a Zeiss Plan-Apochromat objective (40×/1.4 Oil DIC M27), galvanometer-actuated mirrors, and a 780 nm femtosecond laser (Menlo C-Fiber 780 HP). To this end, the MT ferrule is mounted into a dedicated holder. The objective approaches the facets of the fibers in the MT ferrule along the +$y$-direction, see Fig. 1. After immersion in the liquid negative-tone photoresist (IP-Dip, Nanoscribe GmbH), the exact printing positions are found using machine vision. The lithography beam is oriented in parallel to the fiber axes, and the fiber cores are back-illuminated for easier detection.

The FaML on the VCSEL / PD chip are fabricated using a commercially available printing system (Sonata1000, Vanguard Automation GmbH). For simplicity, we print the FaML after mounting the VCSEL / PD arrays to the underlying PCB. In this step, the PCB is fixed by a dedicated holder, and liquid negative-tone photoresist (VanCore B, Vanguard Automation GmbH) is dispensed on the VCSEL / PD arrays. The lithography objective approaches the PCB along the –$z$-direction, see Fig. 1. Again, we use machine-vision to find the printing positions. The axis of the lithography beam is perpendicular to the facets of the VCSEL / PD arrays, and the photoresist serves as an immersion liquid.

Independently of the photoresist and lithography machine used, the exposed structures undergo the same post-processing. After development in propylene-glycol-methyl-etheracetate (PGMEA) for 15 minutes, the samples are flushed with isopropanol, and subsequently blow-dried.

*D. Long-term stability*

Long-term stability of 3D-printed FaML is a key aspect with respect to practical application of the concept. To investigate this aspect, we monitor the evolution of the coupling loss under pertinent damp-heat test conditions. We use a simplified test structure, see Fig. 8, store it in a climate chamber at 85 °C and 85 % relative humidity, and repeatedly measure the optical transmission at a wavelength of $\lambda = 1550\,\text{nm}$ over the course of nearly 4000 hours. The assembly consists of a pair of single mode fibers (SMF) glued into V-grooves. Each of the SMF facets carries a 3D-printed FaML, designed as a loopback: Light coupled into the left SMF enters the first FaML, is redirected by total-internal-reflection (TIR) at the mirror with Surface S0, collimated by the lens Surface S1, and collected by a symmetrically arranged counterpiece, see red beam path in Fig. 8. The FaML consisted of the same photoresist (VanCore B, Vanguard Automation GmbH) used for fabrication of the FaML on the VCSEL / PD chips, see Sections 3 and 4.

We measured the transmission through five identical arrangements as in Fig. 8. The results of these long-term stability tests are shown in Table 1. Within our measurement accuracy, we did not find any sign of degradation for any of the five measured assemblies. The test had to be stopped after 3960 hours, because the single-mode connectors and the coating of the fibers had deteriorated to a degree that would not permit further reliable measurements. The FaML themselves did not show any visible degradation, see Fig. 8.

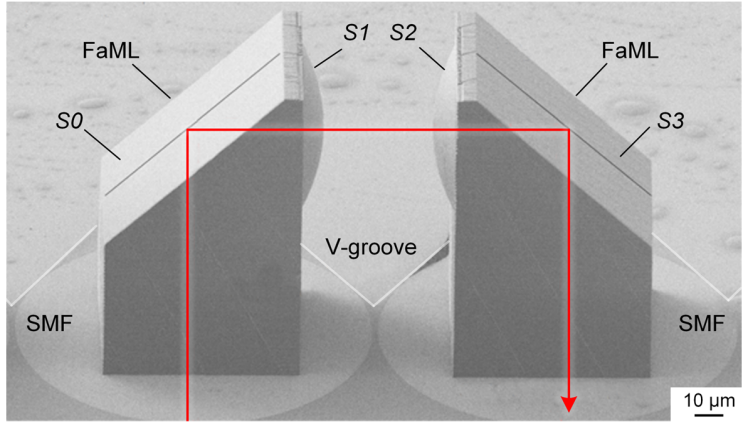

**Fig. 8.** Scanning electron microscopy (SEM) image of a simplified test structure for evaluating the long-term stability of our 3D-printed facet-attached microlenses (FaML). The assembly consists of a pair of single mode fibers (SMF) glued into V-grooves. Each of the SMF facets carries a 3D-printed FaML, designed as a loopback: Light coupled into the left SMF enters the first FaML, is redirected by total-internal-reflection (TIR) at the mirror with Surface S0, and collimated by the lens Surface S1. The beam is then entering the second FaML through the lens Surface S2, is redirected by a second TIR mirror with Surface S3 and finally coupled into the core of the right SMF. The path of the signal is indicated by the red path and arrow. The line in the center of the TIR mirrors has been added for a better orientation.

**Table 1: Long-term stability tests of FaML at a temperature of 85 °C and at a relative humidity of 85 %**

| Connection | Coupling loss [dB] | | | |
|---|---|---|---|---|
| | Initial | 400 h @ 85°C/85% | 1840 h @ 85°C/85% | 3960 h @ 85°C/85% |
| #1 | 1.4 | 1.4 | 1.5 | 1.3 |
| #2 | 1.2 | 1.2 | 1.3 | 1.4 |
| #3 | 1.5 | 1.6 | 1.9 | 1.5 |
| #4 | 1.4 | 1.6 | 1.8 | 1.5 |
| #5 | 1.5 | 1.6 | 1.5 | 1.5 |

In a further set of experiments, we investigated the stability of FaML similar to the ones shown in Fig. 8 at standard reflow-soldering temperatures of up to 260 °C for several minutes. We did not observe any degradation of the measured transmission performance in these experiments.

**Funding.** Deutsche Forschungsgemeinschaft (EXC-2082/1-390761711, CRC 1173); Bundesministerium für Bildung und Forschung (13N14630, 16KISK010, 16ES0948); European Research Council (773248); EU Horizon 2020 Framework Programme (731954); Alfried Krupp von Bohlen und Halbach-Stiftung; Karlsruhe School of Optics and Photonics (KSOP).

**Acknowledgments.** This work was supported by the Deutsche Forschungsgemeinschaft (DFG, German Research Foundation) under Germany's Excellence Strategy via the Excellence Cluster "3D Matter Made to Order" (3DMM2O, EXC-2082/1-390761711) and the Collaborative Research Center WavePhenomena (CRC 1173), by the Bundesministerium für Bildung und Forschung (BMBF) via the projects PRIMA (13N14630), Open6GHub (16KISK010), and DiFeMiS (16ES0948), the latter being part of the programme "Forschungslabore Mikroelektronik Deutschland (ForLab)", by the European Research Council (ERC Consolidator Grant "TeraSHAPE", 773248), by the EU H2020 Photonic Packaging Pilot Line PIXAPP (731954), by the Alfried Krupp von Bohlen und Halbach Foundation, and by the Karlsruhe School of Optics and Photonics (KSOP).

**Disclosures.** P.-I.D. and C.K. are co-founders and shareholders of Vanguard Photonics GmbH and Vanguard Automation GmbH, start-up companies engaged in exploiting 3D nanoprinting in the field of photonic integration and assembly. M.L. and P.-I.D. are employees of Vanguard Automation GmbH. P.M., Y.X., M.T., M.B., P.-I.D., and C.K. are co-inventors of patents owned by Karlsruhe Institute of Technology (KIT) in the technical field of the publication. M.T. and M.B. are now employees of Nanoscribe GmbH, a company selling 3D lithography systems. A.H.-L. and H.K. are employees of Rosenberger Hochfrequenztechnik GmbH & Co. KG, a company selling, e.g., optical connectors for transceiver modules. C.W. is an employee of Rosenberger OSI GmbH & Co. OHG. T.K and A.W. are employees of ficonTEC Service GmbH, a company selling photonic assembly and testing equipment. A.W. is a former employee of Vanguard Automation GmbH. The other authors S.R. and W.F. declare no conflict of interest.

**Data availability.** Data underlying the results presented in this paper are not publicly available at this time but may be obtained from the authors upon reasonable request.